\documentclass[11pt]{article}
\usepackage{amssymb}

\usepackage[totalwidth=460truept,totalheight=600truept]{geometry}
\usepackage{latexsym,amssymb,amsmath,graphicx,dsfont}
\usepackage[T1]{fontenc}

\def\theequation{\arabic{section}.\arabic{equation}}

\renewcommand{\theequation}{\thesection.\arabic{equation}}
\global\arraycolsep=1truept
\begin{document}

\null

\vskip 1.9truecm

\begin{center}
{\huge \textbf{Standard Model Without Elementary Scalars}}

\vskip .3truecm

{\huge \textbf{And High Energy Lorentz Violation}}

\vskip 1.5truecm

\textsl{Damiano Anselmi}

\textit{Dipartimento di Fisica ``Enrico Fermi'', Universit\`{a} di Pisa, }

\textit{Largo Pontecorvo 3, I-56127 Pisa, Italy, }

\textit{and INFN, Sezione di Pisa, Pisa, Italy}

damiano.anselmi@df.unipi.it

\vskip 2truecm

\textbf{Abstract}
\end{center}

\bigskip

{\small If Lorentz symmetry is violated at high energies,  interactions that are usually non-renormalizable can become renormalizable by weighted power counting. Recently, a CPT invariant, Lorentz
violating extension of the Standard Model containing
two scalar-two fermion interactions (which can explain neutrino masses) and four fermion interactions (which can explain proton decay) was proposed. In this paper 
we consider a variant of this model, obtained suppressing the elementary scalar fields, and argue that it can reproduce the known low energy physics.
In the Nambu--Jona-Lasinio spirit, we show, using a large }$N_{c}${\small \
expansion, that a dynamical symmetry breaking takes place. The effective
potential has a Lorentz invariant minimum and the Lorentz violation does not
reverberate down to low energies. The mechanism generates fermion masses,
gauge-boson masses and scalar bound states, to be identified with composite
Higgs bosons. Our approach is not plagued by the ambiguities of approaches
based on non-renormalizable vertices. The low-energy effective action is
uniquely determined and predicts relations among parameters of the Standard
Model.}

\vskip 1truecm

\vfill\eject

\section{Introduction}

\setcounter{equation}{0}

Lorentz symmetry is a basic ingredient of the Standard Model of particle
physics. However, the possibility that it might be violated at very high
energies is still open \cite{colective,kostelecky} and has inspired several
investigations about the new physics that could emerge, at low and high
energies \cite{cohen}.

In quantum field theory, the violation of Lorentz symmetry at high energies
allows us to renormalize otherwise non-renormalizable interactions \cite
{renolor,LVgauge1suA,LVgauge1suAbar}, such as two scalar-two fermion
vertices and four fermion vertices. Terms with higher space derivatives
modify the dispersion relations and generate propagators with improved
ultraviolet behaviors. A ``weighted'' power counting, which assigns
different weights to space and time, allows us to prove that the theory is
renormalizable and consistent with (perturbative) unitarity, namely that no
counterterms with higher time derivatives are generated.

Using these tools, we have recently proposed \cite{LVSM} a Standard Model
extension with the following properties: it is CPT\ invariant, but Lorentz
violating at high energies, it is unitary and renormalizable by weighted
power counting; it contains the vertex $(LH)^{2}/\Lambda _{L}$, which gives
Majorana masses to the neutrinos after symmetry breaking, but no
right-handed neutrinos, nor other extra fields; it contains four fermion
vertices, which can explain proton decay. The scale $\Lambda _{L}\sim
10^{14} $GeV is interpreted as the scale of Lorentz violation. Below that
scale, Lorentz symmetry is recovered.

The model has two ``weighted'' dimensions, which means that at high energies
its power counting resembles the one of a two-dimensional quantum field
theory. In particular, only the four fermion vertices are strictly
renormalizable, while the gauge and Higgs interactions are
super-renormalizable. This means that at energies $\gtrsim \Lambda _{L}$ all
gauge bosons and the Higgs field become free and decouple, and what remains
is a (Lorentz violating) four fermion model in two weighted dimensions. It
is then natural to inquire what physical effects are induced, at lower
energies, by a dynamical symmetry breaking mechanism, in the
Nambu--Jona-Lasinio spirit \cite{njl}. If we suppress the elementary scalar field, we obtain a model that is candidate to 
reproduce the observed low energy physics, predict
relations among otherwise independent parameters, and possibly predict new physics detectable at LHC.

Adapting an old suggestion due to Nambu \cite{nambu}, Miransky \textit{et al}%
. \cite{tcond} and Bardeen \textit{et al}. \cite{tcond2} to our case, we
explore the following scenario. When gauge interactions are switched off,
the dynamical symmetry mechanism produces fermion condensates $\langle \bar{q%
}q\rangle $. The effective potential can be calculated in the large $N_{c}$
limit and has a Lorentz invariant (local) minimum, which gives masses to the
fermions. Massive scalar bound states (composite Higgs bosons) emerge,
together with Goldstone bosons \cite{G}. At a second stage, gauge
interactions are switched back on, so the Goldstone bosons associated with
the breaking of $SU(2)_{L}\times U(1)_{Y}$ to $U(1)_{Q}$ are ``eaten'' by
the $W^{\pm }$ and $Z$ bosons, which become massive.

The low-energy effective action is Lorentz invariant and uniquely
determined. It predicts relations among parameters of the Standard Model.
The naivest predictions are obtained in the leading order of the large $%
N_{c} $ expansion, with gauge interactions switched off, and considering
just the top and bottom quarks. In this simplified situation neutral
composite Higgs bosons have masses $\sim 2m_{t}$ and $\sim 2m_{b}$, and
charged Higgs bosons have masses $\sim \sqrt{2}m_{t}$. The ordinary
single-Higgs situation can be retrieved choosing the four fermion vertices
appropriately, namely squaring the Yukawa coupling to the Higgs field. More
complicate formulas relate $m_{t}$ to the $W$- and $Z$-masses. The leading
order of the $1/N_{c}$ expansion carries a large theoretical error, say
50\%. Curiously, the relation between $m_{t}$ and the Fermi constant turns
out to be in ``too-good'' agreement with the experimental value.

The Nambu--Jona-Lasinio mechanism induces low-energy physics from otherwise
highly suppressed interactions. Since our model is Lorentz violating at high
energies, we can worry that the Lorentz violation might be reverberated down
to low energies. We show that this does not happen, since the minimum of the
effective potential is Lorentz invariant and no Lorentz violating
interactions are drawn down to low energies.

The advantage of our approach with respect to ordinary Nambu--Jona-Lasinio
approaches is that our model is renormalizable, so its high energy behavior
is given, and depends on a certain finite set of free parameters. The
predictivity of non-renormalizable approaches \cite{tcond,tcond2} is
questionable \cite{jmpa}. For example in ref. \cite{hase}, the complete set
of independent Standard Model parameters was generated introducing other
non-renormalizable terms, besides the usual four fermion vertices. Our
model, on the contrary, has an unambiguous high-energy behavior, and can be
used to justify results previously obtained in uncertain theoretical
frameworks. With respect to other approaches to composite Higgs bosons, such
as technicolor \cite{suss}, or the introduction of extra heavy gauge
bosons to renormalize four fermion vertices \cite{jmpa}, it has the
advantage of being conceptually more economic.

The paper is organized as follows. In section 2 we recall the model of ref. 
\cite{LVSM}, present a variant with a simplified gauge sector and introduce
the scalarless model. In section 3 we study the dynamical symmetry breaking
in Lorentz violating four fermion models. In section 4 we study the
phenomenological consequences of this mechanism in our scalarless model, in
particular the generation of fermion masses, bound states, gauge-boson
masses, and so on. In section 5 we study Goldstone's theorem in Lorentz
violating theories. Section 6 contains our conclusions. In Appendix A we
show that a suitable weight rearrangement simplifies the gauge-field sector
of our model (but produces new vertices in the matter sector). In Appendix B
we prove certain mathematical relations that are used in the paper. We work
in Minkoswki spacetime and Wick rotate to Euclidean space when necessary.

\section{The model}

\setcounter{equation}{0}

We assume that invariance under rotations in preserved. We decompose
coordinates $x^{\mu }$ as $(\hat{x}^{\mu },\bar{x}^{\mu })$, where $\hat{x}%
^{\mu }$, or simply $\hat{x}$, denotes the time component (keeping an index
is useful to use the dimensional regularization), and $\bar{x}^{\mu }$
denote the space components. Similarly, we decompose the space time index $%
\mu $ as $(\hat{\mu},\bar{\mu})$, the partial derivative $\partial _{\mu }$
as $(\hat{\partial}_{\mu },\bar{\partial}_{\mu })$, and gauge vectors $%
A_{\mu }$ as $(\hat{A}_{\mu },\bar{A}_{\mu })$. The Lorentz violating theory
is renormalizable by weighted power counting \cite
{renolor,LVgauge1suA,LVgauge1suAbar} in \dj $=1+3/n$ ``weighted
dimensions'', where energy has weight 1 and the space components of momenta
have weight $1/n$. Scalar propagators have weight $-2$ and fermion
propagators have weight $-1$. Details on gauge fields are given in the
Appendix.

The ``Standard-Extended Model'' of ref. \cite{LVSM} has $n=3$ and therefore
weighted dimension 2. The lagrangian of its simplest version reads 
\begin{equation}
\mathcal{L}=\mathcal{L}_{Q}+\mathcal{L}_{\text{kin}f}+\mathcal{L}_{H}+%
\mathcal{L}_{Y}-\frac{\bar{g}^{2}}{4\Lambda _{L}}(LH)^{2}-\sum_{I=1}^{5}%
\frac{1}{\Lambda _{L}^{2}}g\bar{D}\bar{F}\,(\bar{\chi}_{I}\bar{\gamma}\chi
_{I})+\frac{Y_{f}}{\Lambda _{L}^{2}}\bar{\psi}\psi \bar{\psi}\psi -\frac{g}{%
\Lambda _{L}^{2}}\bar{F}^{3},  \label{simplL}
\end{equation}
where 
\begin{eqnarray}
\mathcal{L}_{Q} &=&\frac{1}{4}\sum_{G}\left( 2F_{\hat{\mu}\bar{\nu}}^{G}\eta
^{G}(\bar{\Upsilon})F_{\hat{\mu}\bar{\nu}}^{G}-F_{\bar{\mu}\bar{\nu}%
}^{G}\tau ^{G}(\bar{\Upsilon})F_{\bar{\mu}\bar{\nu}}^{G}\right) ,  \nonumber
\\
\mathcal{L}_{\text{kin}f} &=&\sum_{a,b=1}^{3}\sum_{I=1}^{5}\bar{\chi}_{I}^{a}%
\hspace{0.02in}i\left( \delta ^{ab}\hat{D}\!\!\!\!\slash -\frac{b_{0}^{Iab}}{%
\Lambda _{L}^{2}}{\bar{D}\!\!\!\!\slash}\,^{3}+b_{1}^{Iab}\bar{D}\!\!\!\!%
\slash \right) \chi _{I}^{b},  \nonumber \\
\mathcal{L}_{H} &=&|\hat{D}_{\hat{\mu}}H|^{2}-\frac{a_{0}}{\Lambda _{L}^{4}}|%
\bar{D}^{2}\bar{D}_{\bar{\mu}}H|^{2}-\frac{a_{1}}{\Lambda _{L}^{2}}|\bar{D}%
^{2}H|^{2}-a_{2}|\bar{D}_{\bar{\mu}}H|^{2}-\mu _{H}^{2}|H|^{2}-\frac{\lambda
_{4}\bar{g}^{2}}{4}|H|^{4},  \nonumber \\
\mathcal{L}_{Y} &=&-\bar{g}\Omega _{i}H^{i}+\text{h.c.},\qquad \Omega
_{i}=\sum_{a,b=1}^{3}Y_{1}^{ab}\bar{L}^{ai}\ell _{R}^{b}+Y_{2}^{ab}\bar{u}%
_{R}^{a}Q_{L}^{bj}\varepsilon ^{ji}+Y_{3}^{ab}\bar{Q}_{L}^{ai}d_{R}^{b},
\label{varie}
\end{eqnarray}
$i$,$j$ are $SU(2)_{L}$ indices, $\chi _{1}^{a}=L^{a}=(\nu _{L}^{a},\ell
_{L}^{a})$, $\chi _{2}^{a}=Q_{L}^{a}=(u_{L}^{a},d_{L}^{a})$, $\chi
_{3}^{a}=\ell _{R}^{a}$, $\chi _{4}^{a}=u_{R}^{a}$ and $\chi
_{5}^{a}=d_{R}^{a}$. Moreover, $\nu ^{a}=(\nu _{e},\nu _{\mu },\nu _{\tau })$%
, $\ell ^{a}=(e,\mu ,\tau )$, $u^{a}=(u,c,t)$ and $d^{a}=(d,s,b)$. The sum $%
\sum_{G}$ is over the gauge groups $SU(3)_{c}$, $SU(2)_{L}$ and $U(1)_{Y}$,
and the last three terms of (\ref{simplL}) are symbolic. Finally, $\bar{%
\Upsilon}\equiv -\bar{D}^{2}/\Lambda _{L}^{2}$, where $\Lambda _{L}$ is the
scale of Lorentz violation, and $\eta ^{G}$, $\tau ^{G}$ are polynomials of
degree 2 and 4, respectively. Gauge anomalies cancel out exactly as in the
Standard Model \cite{LVSM}. The ``boundary conditions'' such that Lorentz
invariance is recovered at low energies are that $b_{1}^{Iab}$ tend to $%
\delta ^{ab}$ and $a_{2}$, $\eta ^{G}$ and $\tau ^{G}$ tend to 1 (four such
conditions can be trivially fulfilled normalizing the gauge fields and the
space coordinates $\bar{x}$).

The dispersion relations are modified, because propagators contain higher
powers of the space components of momenta. This improves their ultraviolet
behaviors and makes the theory renormalizable. Since the weight of a scalar
field vanishes in \dj $=2$ a constant $\bar{g}$ of weight $1/2$ is attached
to the scalar legs to ensure renormalizability. The gauge coupling $g$ has
weight 1. The weights of all other parameters are determined so that each
lagrangian term has weight 2 ($=$\dj ). We have neutrino masses $\sim
v^{2}/\Lambda _{L}$, $v$ being the Higgs vev, assuming that all other
parameters involved in the vertex $(LH)^{2}/\Lambda _{L}$ are of order 1.
Reasonable estimates of the neutrino masses (a fraction of eV) give $\Lambda
_{L}\sim 10^{14}$GeV.

An alternative model can be obtained rearranging the weight assignments as
explained in Appendix A, which allows us to simplify the gauge sector.
Specifically, we replace $\eta ^{G}$ with unity and $\tau ^{G}$ with a
polynomial of degree 2, which we denote by $\tau ^{\prime G}$. In a suitable
``Feynman'' gauge the gauge-field propagator becomes reasonably simple to be
used in practical computations. The price is a more complicated Higgs
sector, because $g$ and $\bar{g}$ get a lower weight (1/3). The simplest
version of the alternative model (see the Appendix for details) has
lagrangian 
\begin{eqnarray}
&&\mathcal{L}^{\prime }=\mathcal{L}_{Q}^{\prime }+\mathcal{L}_{\text{kin}f}+%
\mathcal{L}_{H}^{\prime }+\mathcal{L}_{Y}-\frac{\bar{g}^{2}}{4\Lambda _{L}}%
(LH)^{2}-\sum_{I=1}^{5}\frac{1}{\Lambda _{L}^{2}}g\bar{D}\bar{F}\,(\bar{\chi}%
_{I}\bar{\gamma}\chi _{I})+\frac{Y_{f}}{\Lambda _{L}^{2}}\bar{\psi}\psi \bar{%
\psi}\psi -\frac{g}{\Lambda _{L}^{2}}\bar{F}^{3}  \nonumber \\
&&-\frac{1}{\Lambda _{L}^{2}}g\bar{g}\bar{\psi}\psi \bar{F}H-\frac{1}{%
\Lambda _{L}^{2}}\left( \bar{g}^{3}\bar{\psi}\psi H^{3}+\bar{g}^{2}\bar{\psi}%
\bar{D}\psi H^{2}+\bar{g}\bar{\psi}\bar{D}^{2}\psi H\right) -\frac{1}{%
\Lambda _{L}^{4}}\left( g\bar{D}^{2}\bar{F}+g^{2}\bar{F}^{2}\right)
H^{\dagger }H,  \label{simplL2}
\end{eqnarray}
where 
\begin{eqnarray*}
\mathcal{L}_{Q}^{\prime } &=&\frac{1}{4}\sum_{G}\left( 2F_{\hat{\mu}\bar{\nu}%
}^{G}F_{\hat{\mu}\bar{\nu}}^{G}-F_{\bar{\mu}\bar{\nu}}^{G}\tau ^{\prime G}(%
\bar{\Upsilon})F_{\bar{\mu}\bar{\nu}}^{G}\right) , \\
\mathcal{L}_{H}^{\prime } &=&\mathcal{L}_{H}-\frac{\lambda _{4}^{(3)}\bar{g}%
^{2}}{4\Lambda _{L}^{2}}|H|^{2}|\bar{D}_{\bar{\mu}}H|^{2}-\frac{\lambda
_{4}^{(2)}\bar{g}^{2}}{4\Lambda _{L}^{2}}|H^{\dagger }\bar{D}_{\bar{\mu}%
}H|^{2}-\frac{\bar{g}^{2}}{4\Lambda _{L}^{2}}\left[ \lambda
_{4}^{(1)}(H^{\dagger }\bar{D}_{\bar{\mu}}H)^{2}+\text{h.c.}\right] -\frac{%
\lambda _{6}\bar{g}^{4}}{36\Lambda _{L}^{2}}|H|^{6},
\end{eqnarray*}

\paragraph{Scalarless model}

Our scalarless Standard-Extended Model reads 
\begin{equation}
\mathcal{L}_{\mathrm{noH}}=\mathcal{L}_{Q}^{\prime }+\mathcal{L}_{\text{kin}%
f}-\sum_{I=1}^{5}\frac{1}{\Lambda _{L}^{2}}g\bar{D}\bar{F}\,(\bar{\chi}_{I}%
\bar{\gamma}\chi _{I})+\frac{Y_{f}}{\Lambda _{L}^{2}}\bar{\psi}\psi \bar{\psi%
}\psi -\frac{g}{\Lambda _{L}^{2}}\bar{F}^{3},  \label{noH}
\end{equation}
and is obtained suppressing the Higgs field in (\ref{simplL2}). Obviously,
the gauge anomalies of (\ref{noH}) still cancel. We see that the
simplification is considerable.

If we suppress the Higgs field in (\ref{simplL}), the only difference is
that $\mathcal{L}_{Q}$ appears instead of $\mathcal{L}_{Q}^{\prime }$ in (%
\ref{noH}). We keep the simpler model (\ref{noH}), but the conclusions of
this paper do not depend on this choice.

At high energies gauge and Higgs fields become free and decouple, because
their interactions are super-renormalizable, so all theories (\ref{simplL}),
(\ref{simplL2}) and (\ref{noH}) become a four fermion model in two weighted
dimensions, with lagrangian 
\begin{equation}
\mathcal{L}_{\mathrm{4f}}=\sum_{a,b=1}^{3}\sum_{I=1}^{5}\bar{\chi}_{I}^{a}%
\hspace{0.02in}i\left( \delta ^{ab}\hat{\partial}\!\!\!\slash -\frac{%
b_{0}^{Iab}}{\Lambda _{L}^{2}}{\bar{\partial}\!\!\!\slash}\,^{3}+b_{1}^{Iab}%
\bar{\partial}\!\!\!\slash \right) \chi _{I}^{b}+\frac{Y_{f}}{\Lambda
_{L}^{2}}\bar{\psi}\psi \bar{\psi}\psi .  \label{vnoH}
\end{equation}
We have kept also the terms multiplied by $b_{1}^{Iab}$, since they are
necessary to recover Lorentz invariance at low energies.

Our purpose is to investigate whether (\ref{noH}) can describe the known
low-energy physics by means of a dynamical symmetry breaking mechanism
triggered by the four fermion vertices, where some quark-antiquark bilinears
acquire expectation values.

Let us list the candidate condensates. Observe that left- and right-handed
spinors transform in the same way under spatial rotations. Thus, the most
general fermionic bilinears that are scalars under spatial rotations are 
\begin{equation}
(\psi _{1R}^{\dagger }\psi _{2L}),\qquad (\psi _{1L}^{c\dagger }\psi
_{2L}),\qquad (\psi _{1R}^{c\dagger }\psi _{2R}),  \label{f21}
\end{equation}
and their Hermitian conjugates, which are Lorentz invariant, plus 
\begin{equation}
(\psi _{1L}^{\dagger }\psi _{2L}),\qquad (\psi _{1R}^{\dagger }\psi
_{2R}),\qquad (\psi _{1R}^{c\dagger }\psi _{2L}),\qquad (\psi _{1R}^{\dagger
}\psi _{2L}^{c}),  \label{f22}
\end{equation}
which violate both Lorentz symmetry and CPT. We see that every fermion
condensate or mass term that violates Lorentz symmetry violates also CPT.
Thus, at low energies the dynamical symmetry breaking can either preserve
Lorentz symmetry, or break it together with CPT. We show that the effective
potential has a Lorentz invariant minimum.

Consider now the four fermion vertices. The Fierz identity can be used to
convert the structure $(\psi _{1}^{\dagger }\sigma _{i}\psi _{2})(\psi
_{3}^{\dagger }\sigma _{i}\psi _{4})$ into the structure $(\psi
_{1}^{\dagger }\psi _{2}^{\prime })(\psi _{3}^{\dagger }\psi _{4}^{\prime })$%
. Thus, the most general $U(1)_{L}\times U(1)_{R}$-invariant, rotationally
invariant four fermion interactions are 
\begin{equation}
(\psi _{1L}^{\dagger }\psi _{2R})(\psi _{3R}^{\dagger }\psi _{4L}),\qquad
(\psi _{1L}^{\dagger }\psi _{2L})(\psi _{3L}^{\dagger }\psi _{4L}),\qquad
(\psi _{1R}^{\dagger }\psi _{2R})(\psi _{3R}^{\dagger }\psi _{4R}),\qquad
(\psi _{1L}^{\dagger }\psi _{2L})(\psi _{3R}^{\dagger }\psi _{4R}).
\label{str}
\end{equation}
The Lorentz invariant combinations are 
\[
(\psi _{1L}^{\dagger }\psi _{2R})(\psi _{3R}^{\dagger }\psi _{4L}),\qquad
(\psi _{1L}^{\dagger }\psi _{2L}^{c})(\psi _{3L}^{c\dagger }\psi
_{4L}),\qquad (\psi _{1R}^{\dagger }\psi _{2R}^{c})(\psi _{3R}^{c\dagger
}\psi _{4R}),\qquad (\psi _{1L}^{\dagger }\psi _{2R}^{c})(\psi
_{3L}^{\dagger c}\psi _{4R}). 
\]
All combinations are CPT invariant. Lorentz violating four fermion vertices
remain highly suppressed, while the Lorentz invariant ones determine
interactions of the low energy effective theory.

At energy scales much smaller than $\Lambda _{L}$ the low-energy effective
theory resembles a Standard Model with one or more Higgs doublets. However,
the masses of composite Higgs bosons, as well as their self-couplings and
couplings to quarks and gauge fields, are not free, but unambiguously
determined by the model (\ref{noH}).

\paragraph{Predictivity}

The ordinary Nambu--Jona-Lasinio framework \cite{njl} makes use of
non-renormalizable interactions. The dynamical symmetry breaking in
scalarless models was studied in ref.s \cite{tcond,tcond2}. The predictivity
of this approach was questioned in ref. \cite{hase}, where it was shown that
the unknown high-energy physics, duly parametrized, can add enough extra
parameters to the low-energy effective action, and make it completely
equivalent to the Standard Model (with elementary Higgs field), equipped
with all its free constants. The virtue of our approach is that the
high-energy physics of our model is unambiguous, encoded in (\ref{vnoH}).
Since (\ref{vnoH}), as well as (\ref{noH}), (\ref{simplL}) and (\ref{simplL2}%
), are renormalizable by weighted power counting, we do not need to consider
other sectors of unknown physics beyond them. In particular, (\ref{noH})
does not contain the interactions used in \cite{hase} to show the
predictivity loss. Thus, our model is predictive, and actually provides a
viable renormalizable environment for the Nambu--Jona-Lasinio mechanism.

On the other hand, we have a new source of worry. The dynamical symmetry
breaking is a non-perturbative mechanism to generate low-energy effects from
otherwise suppressed high-energy interactions. In our model (\ref{noH}) this
mechanism is triggered by four fermion vertices, which are renormalizable
only thanks to the Lorentz violation. The scale of Lorentz violation $%
\Lambda _{L}$ cannot be treated as a cut-off in an otherwise renormalizable
Lorentz invariant theory (in that case, it would be possible to completely
recover Lorentz symmetry at low energies in an obvious way). The dynamical
symmetry breaking might reverberate the Lorentz violation down to low
energies. If that happened, our scalarless model (\ref{noH}) would be in
trouble. One of our goals is to prove that the violation of Lorentz symmetry
remains highly suppressed even when the dynamical symmetry breaking takes
place. Crucial for the proof is the fact, noted above, that Lorentz
violating fermion condensates violate also CPT. In some sense, this raises
the price of low-energy Lorentz violation enough to disfavour it.

\paragraph{CPT}

In this paper, as in \cite{LVSM}, we assume exact CPT\ invariance. While (in
local, Hermitian) theories a CPT violation implies also the violation of
Lorentz symmetry, the converse is not true, in general, except for special
subclasses of terms, such as the fermionic bilinears (\ref{f21}) and (\ref
{f22}). Thus, we have to introduce two a priori different energy scales, a
scale of Lorentz violation $\Lambda _{L}$, and a scale of CPT violation $%
\Lambda _{\mathrm{CPT}}$, with $\Lambda _{\mathrm{CPT}}\geqslant \Lambda
_{L} $. Our estimate $\Lambda _{L}\sim 10^{14}$GeV is obtained without using
the recent bounds on Lorentz violation suggested by the analysis of $\gamma $%
-ray bursts \cite{ellis}, which claim that the first correction 
\begin{equation}
c(E)\sim c\left( 1-\frac{E}{\bar{M}}\right)  \label{ell}
\end{equation}
to the velocity of light involves an energy scale $\bar{M}\geqslant 1.3\cdot
10^{18}$GeV. In the realm of local perturbative quantum field theory, a
dispersion relation giving (\ref{ell}) must contain odd powers of the
energy, therefore it must also violate CPT. Thus, we are lead to interpret
the results of \cite{ellis} as bounds on $\Lambda _{\mathrm{CPT}}=\bar{M}$
rather than $\Lambda _{L}$. It is conceivable that there exists an energy
region $\Lambda _{L}\leqslant E\leqslant \Lambda _{\mathrm{CPT}}$ where
Lorentz symmetry is violated but CPT is still conserved. Assuming $\Lambda _{%
\mathrm{CPT}}\geqslant M_{Pl}$, we expect that this region spans at least
four or five orders of magnitude. This argument justifies our assumption of
CPT invariance.

\section{Dynamical symmetry breaking}

\setcounter{equation}{0}

In this section we illustrate the dynamical symmetry breaking in a simple
Lorentz violating four fermion model. We show that there exists a Lorentz
invariant minimum, and that the Lorentz violation remains highly suppressed.
In the next section we derive phenomenological consequences for the
scalarless model (\ref{noH}).

We consider the model 
\begin{equation}
\mathcal{L}_{q}=\sum_{I=1}^{N}\bar{\psi}_{I}\left( \gamma ^{\mu }i\left( 
\hat{\partial}_{\mu }+{\bar{\partial}}_{\mu }-{\bar{\partial}}_{\mu }\frac{{%
\bar{\partial}}^{2}}{\Lambda _{L}^{2}}\right) -M\right) \psi _{I}-V_{2}(M),
\label{qm}
\end{equation}
in the leading order of the large $N$ expansion. We have introduced real
auxiliary fields $\rho _{\pm }$ and a complex auxiliary field $\tau $, such
that, in the basis $\psi =(\psi _{L},\psi _{R})$, 
\begin{equation}
M=\left( 
\begin{tabular}{cc}
$\tau $ & $\rho _{+}-\rho _{-}$ \\ 
$\rho _{+}+\rho _{-}$ & $\bar{\tau}$%
\end{tabular}
\right) ,  \label{nota}
\end{equation}
and 
\[
V_{2}(M)=\frac{\Lambda _{L}^{2}}{\lambda ^{2}}|\tau |^{2}+\frac{\Lambda
_{L}^{2}}{2g_{+}^{2}}\rho _{+}^{2}+\frac{\Lambda _{L}^{2}}{2g_{-}^{2}}\rho
_{-}^{2}+\frac{\Lambda _{L}^{2}}{g_{+-}^{2}}\rho _{+}\rho _{-}. 
\]
The four fermion vertices are obtained integrating $\rho _{\pm }$ and $\tau $
out. We keep only the combinations of the form $(\psi _{1I}^{\dagger }\psi
_{2I})(\psi _{3J}^{\dagger }\psi _{4J})$, which contribute to scalar
intermediate channels in the leading order.

We could introduce also parameters $b_{0R}$, $b_{0L}$ and $b_{1R}$, $b_{1L}$
in front of ${\bar{\partial}\!\!\!\slash}\,^{3}$ and $\bar{\partial}\!\!\!%
\slash $, as in (\ref{vnoH}). However, the $\psi $ self-energy receives no
renormalization to the lowest order. Thus, in our approximation $b_{1R}$ and 
$b_{1L}$ must be set equal to 1, to have Lorentz invariance at low energies.
We also make the simplifying assumption $b_{0R}=b_{0L}\equiv b_{0}$ and
reabsorb $b_{0}$ inside $\Lambda _{L}$. The model (\ref{qm}) is
renormalizable as it stands in the leading order.

A non-vanishing $\tau $ vacuum expectation value gives the fermions a Dirac
mass. On the other hand, non-trivial $\rho _{\pm }$ expectation values
correspond to Lorentz and CPT\ violating mass terms of the form $\bar{\psi}%
_{I}\gamma ^{0}\psi _{I}$ and $\bar{\psi}_{I}\gamma ^{0}\gamma _{5}\psi _{I}$%
.

The $M$-effective potential $V(M)$ can be calculated assuming that $\rho
_{\pm }$ and $\tau $ are constants. The leading contributions come from
fermion loops with $\rho _{\pm }$ and $\tau $ external legs. We have 
\[
V(M)=V_{2}(M)+iN\int \frac{\text{\textrm{d}}^{4}p}{(2\pi )^{4}}\ln \det
(-\gamma ^{\mu }p_{\mu }^{\prime }+M), 
\]
having defined 
\begin{equation}
\hat{p}^{\prime \hspace{0.01in}\mu }=\hat{p}^{\mu },\qquad \bar{p}^{\prime 
\hspace{0.01in}\mu }=\bar{p}^{\mu }\left( \frac{\bar{p}^{2}}{\Lambda _{L}^{2}%
}+1\right) .  \label{uy}
\end{equation}
Using invariance under rotations, we can orient $\bar{p}^{\prime \hspace{%
0.01in}\mu }$ along the $z$ direction. Then we find 
\[
\det \left( -\gamma ^{\mu }p_{\mu }^{\prime }+M\right) =T_{+}T_{-},\qquad
T_{\pm }\equiv |\tau |^{2}-(\hat{p}-\rho _{+})^{2}+(|\bar{p}^{\prime }|\pm
\rho _{-})^{2}. 
\]
Splitting the integral as the sum of two integrals, one for each
contribution $T_{\pm }$, and translating $\hat{p}$, we find, after the Wick
rotation, 
\[
V(M)=V_{2}(M)+v(\rho _{-},\tau )+v(-\rho _{-},\tau ),\qquad v(\rho ,\tau
)=-N\int^{\Lambda }\frac{\text{\textrm{d}}^{4}p}{(2\pi )^{4}}\ln \left(
|\tau |^{2}+\hat{p}^{2}+(|\bar{p}^{\prime }|+\rho )^{2}\right) . 
\]
The integral is divergent and regulated with a cut-off $\Lambda \gg \Lambda
_{L}$. Observe that the corrections to $V_{2}(M)$ are $\rho _{+}$%
-independent.

We first work at $\rho _{-}=0$, find the tentative minimum and later prove
that it does remain a minimum once $\rho _{-}$ is switched on. Rescaling the
momentum $p$ to $p\Lambda _{L}$ (and the cut-off $\Lambda $ to $\Lambda
/\Lambda _{L}$) and defining $|\sigma |^{2}=|\tau |^{2}/\Lambda _{L}^{2}$,
we find, up to an irrelevant additive constant, 
\[
V(0,\tau )=\Lambda _{L}^{2}\frac{|\tau |^{2}}{\lambda ^{2}}+2N\Lambda
_{L}^{4}v(|\sigma |^{2}), 
\]
where the function $v$ is defined in Appendix B, formula (\ref{d1}).
Differentiating once with respect to $\tau $, using (\ref{g3}) and
subtracting the logarithmic divergence (which amounts to replace the cut-off 
$\Lambda $ with the dynamical scale $\mu $), we obtain the \textit{gap
equation} 
\begin{equation}
\Lambda _{\mathrm{RG}}^{3}=\frac{\Lambda _{L}^{2}}{2}\sqrt{\langle |\tau
|^{2}\rangle }\exp (12\pi ^{2}\bar{v}^{\prime }(\langle |\sigma |^{2}\rangle
))>0,\qquad \Lambda _{\mathrm{RG}}=\mu \exp \left( -\frac{2\pi ^{2}}{\lambda
^{2}N}\right) ,  \label{gap}
\end{equation}
for the non-trivial vacuum expectation value $\langle |\sigma |^{2}\rangle $%
, where $\bar{v}^{\prime }$ is the finite function defined in (\ref{g2}).
Since $\Lambda _{\mathrm{RG}}$ is a free parameter, $\langle |\sigma
|^{2}\rangle $ cannot be determined, and equation (\ref{gap}) just relates $%
\langle |\sigma |^{2}\rangle $ and $\Lambda _{\mathrm{RG}}$. Choosing $%
\Lambda _{\mathrm{RG}}$ appropriately, the gap equation has any solution $%
\langle |\sigma |^{2}\rangle $ we like. This arbitrariness is going to
disappear from every other physical quantity. Observe that the $\tau $%
-sector is asymptotically free in the large $N$ expansion.

The ratio $\langle |\sigma |\rangle =\langle |\tau |\rangle /\Lambda _{L}$
between the fermion mass $\langle |\tau |\rangle $ and the scale of Lorentz
violation is very small. Typical values are $\langle |\tau |\rangle \sim
m_{t}\sim 171\mathrm{GeV}$ and $\Lambda _{L}\sim 10^{14}\mathrm{GeV}$, so $%
\langle |\sigma |\rangle \sim 10^{-12}$. Thus, it is meaningful to expand
for $\langle |\sigma |^{2}\rangle \ll 1$, which can be done with the help of
formula (\ref{g1}). The gap equation becomes 
\[
\frac{\Lambda _{\mathrm{RG}}^{2}}{\Lambda _{L}^{2}}-1\sim \frac{1}{2}\frac{%
\langle |\tau |^{2}\rangle }{\Lambda _{L}^{2}}\ln \frac{\Lambda _{L}^{2}}{%
\langle |\tau |^{2}\rangle }\sim 10^{-22}, 
\]
which exhibits a fine-tuning problem, associated with the quadratic
divergences arising for large $\Lambda _{L}$. Nevertheless, this problem is
isolated to the gap equation, since all other quantities we are going to
work with depend on $\Lambda _{L}$ only logarithmically.

Expanding around $|\tau |^{2}=\langle |\tau |^{2}\rangle $, the potential at
vanishing $\rho _{\pm }$ reads 
\begin{equation}
V(0,\tau )\sim 2N|\tau -\langle \tau \rangle |^{2}\langle |\tau |^{2}\rangle
v^{\prime \prime }(\langle |\sigma |^{2}\rangle ),  \label{poto}
\end{equation}
which is finite and positive (see (\ref{2d})). This proves that $|\tau
|^{2}=\langle |\tau |^{2}\rangle $ is indeed a minimum at $\rho _{\pm }=0$.
For $\langle |\sigma |^{2}\rangle \ll 1$ we have, using (\ref{v2}), 
\[
V(0,\tau )\sim |\tau -\langle \tau \rangle |^{2}\frac{N\langle |\tau
|^{2}\rangle }{8\pi ^{2}}\ln \frac{\Lambda _{L}^{2}}{\langle |\tau
|^{2}\rangle }. 
\]

Finally, we switch the fields $\rho $ back on and prove that $|\tau
|^{2}=\langle |\tau |^{2}\rangle $, $\rho _{\pm }=0$ remains a minimum. The
first derivative of $V(\rho ,\tau )$ with respect to $\rho $, calculated at $%
\rho =0$, must necessarily vanish, since $\rho $ is CPT\ odd. The same is
true for the second derivatives with respect to $\rho $ and $\tau $. On the
other hand, the second derivative of $v(\rho ,\tau )$ with respect to $\rho $%
, calculated at $\rho =0$, is finite and $g$-independent (and actually,
``small''). Precisely, we find, using (\ref{an1}), 
\begin{equation}
\left. \frac{\partial ^{2}v}{\partial \rho ^{2}}\right| _{\min }=-4N\langle
|\tau |^{2}\rangle v^{\prime \prime }(\langle |\sigma |^{2}\rangle )<0.
\label{rmi}
\end{equation}
Although negative, this is a finite quantity, independent of $g_{-}$, so it
can always be beaten by $V_{2}(M)$, choosing the coupling $g_{-}$
sufficiently small. Under this assumption, $|\tau |^{2}=\langle |\tau
|^{2}\rangle $, $\rho _{\pm }=0$ does remain a minimum after switching $\rho
_{\pm }$ on.

Using (\ref{v2}) again, the effective potential around the minimum reads for 
$\langle |\tau |^{2}\rangle \ll \Lambda _{L}^{2}$, 
\begin{equation}
V(\rho ,\tau )\sim \frac{\Lambda _{L}^{2}}{2g_{+}^{2}}\rho _{+}^{2}+\frac{%
\Lambda _{L}^{2}}{2g_{-}^{2}}\rho _{-}^{2}\left( 1-\frac{g_{-}^{2}N}{2\pi
^{2}}\frac{\langle |\tau |^{2}\rangle }{\Lambda _{L}^{2}}\ln \frac{\Lambda
_{L}^{2}}{\langle |\tau |^{2}\rangle }\right) +\frac{\Lambda _{L}^{2}}{%
g_{+-}^{2}}\rho _{+}\rho _{-}+|\tau -\langle \tau \rangle |^{2}\frac{%
N\langle |\tau |^{2}\rangle }{8\pi ^{2}}\ln \frac{\Lambda _{L}^{2}}{\langle
|\tau |^{2}\rangle }.  \label{vapp}
\end{equation}
We see that the coefficient of $\rho _{-}^{2}$ receives a very small
correction, of the order $10^{-23}$, even taking $g_{L}^{2}N$ and $%
g_{-}^{2}N $ of order 1. The Lorentz invariant minimum could be spoiled only
if $g_{-}^{2}N$ had an inordinate value.

Before proceeding, a comment is in order. We have so far worked in the large 
$N$ expansion, concentrating on the leading order. In the case of the
Standard Model, we are going to expand for large number of colors $N_{c}$ 
\cite{tcond2}. The reason is that, because of the intrinsically
non-perturbative nature of minima such as the one encoded in (\ref{gap}), we
have control on them only in an expansion of this type. For example, (\ref
{gap}) implies, using (\ref{g1}), 
\[
-\frac{1}{2\lambda ^{2}N}=v^{\prime }(\langle |\sigma |^{2}\rangle )=-\frac{1%
}{4\pi ^{2}}\ln \mu +\cdots . 
\]
The left-hand side of this expression is singular in the ordinary
perturbative expansion ($\lambda \ll 1$), but regular in the large $N$
expansion ($\lambda N^{2}\sim 1$). At finite $N$, higher order corrections
contain arbitrarily high powers of $\lambda ^{2}\ln \mu $ (``leading
logs''), plus powers of $\lambda ^{2}\ln \mu $ multiplied by one extra
factor of $\lambda ^{2}$ (``next-to-leading logs''), and so on. The
resummation of leading logs involves only one-loop results, and can be
easily done. Useful references for such kind of resummations applied to the
Coleman-Weinberg mechanism are for example \cite{sher,iso}. However, it is
known \cite{sher} that in general the so-corrected potential does not
exhibit the nice features of the one calculated in the large $N$ expansion.
Most of the times the minimum suggested by a one-loop truncation (which
cannot be trusted unless it is combined with a large $N$ expansion or a
dimensional transmutation \cite{CW}) is spoiled by the resummation,
depending on the model. Although we can generically expect that the exact
potential will have a non-trivial minimum in most theories, here we want to
have some explicit control on the vacuum, for example check that it
preserves Lorentz symmetry. At present we can answer our questions only in
the large $N$ expansion.

\section{Masses and bound states in scalarless model}

\setcounter{equation}{0}

In this section we show how the masses of quarks and gauge bosons, as well
as composite Higgs bosons, emerge in the scalarless model (\ref{noH}). We
start with the $t$-$b$ model

\begin{equation}
\mathcal{L}_{q}=\sum_{I=1}^{N_{c}}\bar{\psi}_{I}\left( \Gamma ^{\mu }i\left( 
\hat{\partial}_{\mu }+{\bar{\partial}}_{\mu }-{\bar{\partial}}_{\mu }\frac{{%
\bar{\partial}}^{2}}{\Lambda _{L}^{2}}\right) -M\right) \psi _{I}-V_{2}(M),
\label{model2}
\end{equation}
where 
\[
M=\left( 
\begin{tabular}{cc}
$\tau $ & $\rho _{R}$ \\ 
$\rho _{L}$ & $\tau ^{\dagger }$%
\end{tabular}
\right) ,\quad \psi =\left( 
\begin{tabular}{c}
$Q_{L}^{i}$ \\ 
$Q_{R}^{k}$%
\end{tabular}
\right) , 
\]
$Q=(t,b)$, $i,j,\ldots $ are indices of $SU(2)_{L}$ or $SU(2)_{R}$,
depending on the case, $(\Gamma ^{\mu })_{\alpha \beta }^{ij}=\gamma
_{\alpha \beta }^{\mu }\delta ^{ij}$, $\tau $, $\rho _{R}$ and $\rho _{L}$
are 2$\times $2 matrices of fields, and $\rho _{R}$, $\rho _{L}$ are
Hermitian. The most general $SU(2)_{L}\times U(1)_{Y}\times U(1)_{B}\times
U(1)_{A}$-invariant quadratic potential is 
\begin{equation}
\Lambda _{L}^{-2}V_{2}(M)=\mathrm{tr}[\tau \tau ^{\dagger }C]+\frac{1}{%
2g_{L}^{2}}\mathrm{tr}[\rho _{L}^{2}]+\frac{1}{2g_{L}^{\prime \hspace{0.01in}%
2}}(\mathrm{tr}[\rho _{L}])^{2}+g_{R}^{kl}\mathrm{tr}[\rho _{L}]\rho
_{R}^{kl}+\frac{1}{2}g_{RR}^{klmn}\rho _{R}^{kl}\rho _{R}^{mn},  \label{v2m}
\end{equation}
where $C_{ij}$, $g_{L}$, $g_{L}^{\prime }$, $g_{R}^{kl}$ and $g_{RR}^{klmn}$
are constants, $C_{ij}$ and $g_{R}^{kl}$ are diagonal and $g_{RR}^{klmn}$
are non-vanishing only for $k=l,m=n$ and $k=n,l=m$. Although we do not
assume any ``custodial'' $SU(2)_{R}$-invariance, which is indeed violated by 
$V_{2}(M)$, note that $\mathcal{L}_{q}+V_{2}(M)$ is invariant under $%
U(2)_{L}\times U(2)_{R}$ (transforming $\tau $, $\rho _{R}$ and $\rho _{L}$
appropriately), therefore so is the one-loop correction to the effective
potential.

We want to prove that 
\begin{equation}
\tau =\left( 
\begin{tabular}{cc}
$m_{t}$ & $0$ \\ 
$0$ & $m_{b}$%
\end{tabular}
\right) \equiv \tau _{0},\qquad \rho _{L}=\rho _{R}=0.  \label{mini}
\end{equation}
is a minimum, where $m_{t}>m_{b}$ can be identified with the top and bottom
masses, respectively, and are related to the $C$ entries (see below). The
vacuum (\ref{mini}) breaks $SU(2)_{L}\times U(1)_{Y}\times U(1)_{B}\times
U(1)_{A}$ to $U(1)_{Q}\times U(1)_{B}$.

Again, we first work at $\rho _{L}=\rho _{R}=0$, find the tentative minimum
of the effective potential and later prove that it remains a minimum when $%
\rho _{L}$ and $\rho _{R}$ are switched back on. At $\rho _{R}=\rho _{L}=0$
the determinant of $-\gamma ^{\mu }p_{\mu }^{\prime }+M$ is a Lorentz
invariant polynomial of the four-vector $p_{\mu }^{\prime }$ and can be
easily calculated first at $\bar{p}_{\mu }^{\prime }=0$, then replacing $%
\hat{p}^{2}$ with $p_{\mu }^{\prime }p^{\prime \hspace{0.01in}\mu }$. We
find 
\[
\det (-\gamma ^{\mu }p_{\mu }^{\prime }+M)=\left[ (p^{\prime \hspace{0.01in}%
2})^{2}-p^{\prime \hspace{0.01in}2}t_{1}+\frac{1}{2}(t_{1}^{2}-t_{2})\right]
^{2},\qquad t_{1}=\mathrm{tr}[\tau \tau ^{\dagger }],\qquad t_{2}=\mathrm{tr}%
[\tau \tau ^{\dagger }\tau \tau ^{\dagger }]. 
\]
It is easy to prove the inequalities 
\[
t_{1}^{2}\geqslant t_{2}\geqslant \frac{1}{2}t_{1}^{2}. 
\]
The second inequality follows from $\mathrm{tr}[\mathrm{N}^{2}]\geqslant 
\mathrm{tr}[\mathrm{N}]^{2}/2$, where $\mathrm{N}$ is any Hermitian $2\times
2$ matrix. We find the one-loop effective potential 
\[
V(\tau )=\Lambda _{L}^{2}\mathrm{tr}[\tau \tau ^{\dagger }C]+2N_{c}\Lambda
_{L}^{4}\left( v(r_{+})+v(r_{-})\right) , 
\]
where 
\[
r_{\pm }=\frac{1}{2\Lambda _{L}^{2}}\left( t_{1}\pm \sqrt{2t_{2}-t_{1}^{2}}%
\right) \geqslant 0, 
\]
and $v(r)$ is defined in formula (\ref{d1}). The minimum (\ref{mini}) has $%
r_{+}=m_{t}^{2}/\Lambda _{L}^{2}\equiv r_{t}$, $r_{-}=m_{b}^{2}/\Lambda
_{L}^{2}\equiv r_{b}$. Taylor-expanding $v(r)$ around (\ref{mini}), we can
write, neglecting additive constants, 
\begin{eqnarray*}
V(\tau ) &=&\Lambda _{L}^{2}V_{a}(\tau )+\Lambda _{L}^{4}V_{b}(\tau ), \\
V_{a}(\tau ) &=&\mathrm{tr}[\tau \tau ^{\dagger }C]+2\Lambda
_{L}^{2}N_{c}\left( r_{+}v^{\prime }(r_{t})+r_{-}v^{\prime }(r_{b})\right) ,
\\
V_{b}(\tau ) &=&N_{c}(r_{+}-r_{t})^{2}v^{\prime \prime
}(r_{t})+N_{c}(r_{-}-r_{b})^{2}v^{\prime \prime }(r_{b}).
\end{eqnarray*}
Setting the first derivatives of $V_{a}(\tau )$ to zero gives the gap
equations 
\begin{equation}
c_{t}=-2N_{c}v^{\prime }(r_{t}),\qquad c_{b}=-2N_{c}v^{\prime }(r_{b}),
\label{gap2}
\end{equation}
where $C=\mathrm{diag}(c_{t},c_{b})$. Since $c_{t}$ and $c_{b}$ are free
parameters, they can always be chosen so that the gap equations have
solutions. Now, (\ref{mini}) is a minimum of $V_{b}(\tau )$, because $%
v^{\prime \prime }(r)>0$. Moreover, expanding $V_{a}(\tau )$ around (\ref
{mini}), we find 
\[
V_{a}(\tau )\sim 2N_{c}\frac{v^{\prime }(r_{t})-v^{\prime }(r_{b})}{%
m_{t}^{2}-m_{b}^{2}}|m_{t}\tau _{21}+m_{b}\bar{\tau}_{12}|^{2}\geqslant 0. 
\]
The coefficient is positive because of (\ref{rinc}). Thus, (\ref{mini}) is a
minimum of the effective potential at $\rho _{L}=\rho _{R}=0$. There are of
course flat directions $m_{t}\delta \tau _{21}+m_{b}\delta \bar{\tau}_{12}=0$
corresponding to the charged Goldstone bosons (see below).

When $\rho _{L}$ and $\rho _{R}$ are switched on, we can proceed as in the
previous section. The first derivatives of the effective potential around (%
\ref{mini}) still vanish, by CPT invariance, as well as the second
derivatives with respect to one $\tau $-entry and one $\rho $-entry. On the
other hand, the second derivatives with respect to $\rho $-entries are
finite, and can always be made positive choosing the arbitrary constants $%
g_{L}$, $g_{L}^{\prime }$, $g_{R}^{kl}$ and $g_{RR}^{klmn}$ in (\ref{v2m})
appropriately. The reason why the second $\rho $-derivatives have no UV
divergences is that in the corresponding integrals $\gamma ^{\mu }p_{\mu
}^{\prime }$ is sandwiched between two $\gamma ^{0}$'s. Then, using (\ref
{ant}) with $k=2$, we have, in Euclidean space, 
\begin{eqnarray}
&&\int \frac{\mathrm{d}^{4}p}{(2\pi )^{4}}\frac{\gamma ^{0}\gamma ^{\mu
}p_{\mu }^{\prime }\gamma ^{0}\gamma ^{\nu }p_{\nu }^{\prime }}{(p^{\prime 
\hspace{0.01in}2}+m_{1}^{2})(p^{\prime \hspace{0.01in}2}+m_{2}^{2})}=\int 
\frac{\mathrm{d}^{4}p}{(2\pi )^{4}}\frac{\hat{p}^{2}-\bar{p}^{\prime \hspace{%
0.01in}2}-2\hat{p}\gamma ^{0}\gamma ^{\bar{\nu}}\bar{p}_{\bar{\nu}}^{\prime }%
}{(p^{\prime \hspace{0.01in}2}+m_{1}^{2})(p^{\prime \hspace{0.01in}%
2}+m_{2}^{2})}  \nonumber \\
&&\qquad =\int_{0}^{1}\mathrm{d}x\hspace{0.01in}\left[
m_{1}^{2}x+m_{2}^{2}(1-x)\right] \int \frac{\mathrm{d}^{4}p}{(2\pi )^{4}}%
\frac{1}{(p^{\prime \hspace{0.01in}2}+m_{1}^{2}x+m_{2}^{2}(1-x))^{2}}<\infty
.  \label{info}
\end{eqnarray}

Thus, we have proved that (\ref{mini}) is indeed a minimum of the effective
potential. Because of the arbitrariness of $c_{t}$ and $c_{b}$, the top and
bottom masses cannot be predicted. However, they can be related to other
known quantities.

\paragraph{Bound states}

We write $\tau =\tau _{0}+\eta $. The contributions $\Gamma _{\eta \eta }$
and $\Gamma _{\rho \rho }$ to the generating functional $\Gamma $ that are
quadratic in $\eta $ and $\rho $ give the dynamically generated propagators
of such fields, from which we can read the bound states. We expect four
massless scalars in the $\tau $-sector, which are the Goldstone bosons
associated with the broken generators of $SU(2)_{L}\times U(1)_{Y}$ and $%
U(1)_{A}$. Using the solutions (\ref{gap2}) to the gap equations, every $%
\Lambda $-divergences cancel out and both $\Gamma _{\eta \eta }$ and $\Gamma
_{\rho \rho }$ are given by finite integrals. If we are interested in the
low-energy limit with respect to $\Lambda _{L}$, then we can view our
(finite, but Lorentz violating) integrals as usual Lorentz invariant
integrals regulated by the cut-off $\Lambda _{L}$. Such a cut-off is Lorentz
violating, but invariant under translations, spatial rotations and CPT.
Using power counting, it can be easily checked that the $\Lambda _{L}$%
-divergences are linear or logarithmic. However, linear divergences are
absent by CPT and rotational invariance. On the other hand, logarithmic
divergences do not depend on the regulator. In particular, they are Lorentz
invariant, as are the finite parts. Thus, to study the large $\Lambda _{L}$
limit we can regulate our integrals in the most convenient way, e.g.
integrating suitable Lorentz invariant integrands over momenta $p\leqslant
\Lambda _{L}$. Specifically, we can perform the replacement 
\[
\int \frac{\mathrm{d}^{4}p}{(2\pi )^{4}}f(p,\Lambda _{L})\rightarrow
\int^{\Lambda _{L}}\frac{\mathrm{d}^{4}p}{(2\pi )^{4}}f(p,\infty ) 
\]
and use a symmetric integration to kill linear divergences.

We first calculate the leading contributions to $\Gamma _{\eta \eta }$.
Using the tricks mentioned above, we find, in momentum space, 
\[
\Gamma _{\eta \eta }=N_{c}\sum_{i,j=t,b}\left\{ 2\eta _{ij}(p)\bar{\eta}%
_{ij}(-p)(p^{2}f_{ij}^{\prime }-m_{j}^{2}f_{ij})-m_{i}m_{j}f_{ij}\left[ \eta
_{ij}(p)\eta _{ji}(-p)+\bar{\eta}_{ij}(p)\bar{\eta}_{ji}(-p)\right] \right\}
, 
\]
the functions $f_{ij}(p^{2})$ and $f_{ij}^{\prime }(p^{2})$ being defined in
(\ref{ff}). Studying the poles of the effective $\eta $-propagators we find:

\noindent 1) two neutral massive bound states of squared masses 
\[
2m_{t}^{2}\frac{f_{tt}}{f_{tt}^{\prime }}=4m_{t}^{2},\qquad 2m_{b}^{2}\frac{%
f_{bb}}{f_{bb}^{\prime }}=4m_{b}^{2}; 
\]

\noindent 2) two neutral and two charged massless states, which are the
Goldstone bosons;

\noindent 3) two charged massive bound states of squared masses 
\[
f_{tb}\left( \frac{m_{t}^{2}}{f_{bt}^{\prime }}+\frac{m_{b}^{2}}{%
f_{tb}^{\prime }}\right) \sim 2m_{t}^{2}. 
\]
We have used (\ref{appro}) for the approximate values.

The $\rho $--self-energies do not give bound states at low energies. Using
tricks similar to those leading to (\ref{info}) we find 
\begin{eqnarray*}
\Gamma _{\rho \rho } &=&-\frac{\Lambda _{L}^{2}}{2g_{L}^{2}}\mathrm{tr}[\rho
_{L}^{2}]-\frac{\Lambda _{L}^{2}}{2g_{L}^{\prime \hspace{0.01in}2}}(\mathrm{%
tr}[\rho _{L}])^{2}-\Lambda _{L}^{2}g_{R}^{kl}\mathrm{tr}[\rho _{L}]\rho
_{R}^{kl}-\frac{\Lambda _{L}^{2}}{2}g_{RR}^{klmn}\rho _{R}^{kl}\rho
_{R}^{mn}+ \\
&&+N_{c}\sum_{i,j=t,b}\left[ \left( \rho _{L}^{ij}\rho _{L}^{ji}+\rho
_{R}^{ij}\rho _{R}^{ji}\right) \left( 2\mathbf{p}^{2}f_{ij}^{\prime \prime
}+m_{i}^{2}f_{ij}^{\prime }+m_{j}^{2}f_{ji}^{\prime }\right) -2\rho
_{L}^{ij}\rho _{R}^{ji}m_{i}m_{j}f_{ij}\right] .
\end{eqnarray*}
The spatial squared momentum $\mathbf{p}^{2}$ appears instead of $p^{2}$,
which signals the absence of bound states. Moreover, no gap equation
reabsorbs the $\mathcal{O}(\Lambda _{L}^{2})$-terms of $V_{2}(M)$. This
means that all corrections to $V_{2}(M)$ are negligible at low energies, so
even if some $\rho $-bound state existed, it would have masses of the order $%
\sim \Lambda _{L}$. We conclude that the Lorentz violation does not
reverberate down to low energies.

\paragraph{Masses of the gauge bosons}

As usual, when gauge interactions are switched on the three Goldstone bosons 
$\phi ^{\pm }$ and $\phi ^{0}$ associated with the broken generators of $%
SU(2)_{L}\times U(1)_{Y}$ are ``eaten'' by the gauge bosons $W^{\pm }$ and $%
Z $, which become massive. To see how that happens in our case, we proceed
as follows. We first calculate the leading contributions to the $W$-$\eta $, 
$Z$-$\eta $ and $A$-$\eta $ two-point functions. They are given by the
diagrams constructed with one fermion loop, one vertex $\bar{\psi}M\psi $
and one vertex 
\begin{equation}
g(W_{\mu }^{+}J_{+}^{\mu }+W_{\mu }^{-}J_{-}^{\mu })+\tilde{g}Z_{\mu
}J_{Z}^{\mu }+eA_{\mu }J_{\mathrm{em}}^{\mu },  \label{nu}
\end{equation}
where $\tilde{g}=\sqrt{g^{2}+g^{\prime 2}}$. We find 
\begin{equation}
\Gamma _{A\eta }=-gN_{c}(\partial ^{\mu }\phi ^{-})W_{\mu
}^{+}-gN_{c}(\partial ^{\mu }\phi ^{+})W_{\mu }^{-}-N_{c}\tilde{g}(\partial
^{\mu }\phi ^{0})Z_{\mu },  \label{gat}
\end{equation}
where 
\[
\phi ^{+}=i\sqrt{2}(m_{t}f_{tb}^{\prime }\eta _{tb}-m_{b}f_{bt}^{\prime }%
\bar{\eta}_{bt}),\qquad \phi ^{0}=\frac{i}{2}(m_{t}f_{tt}(\eta _{tt}-\bar{%
\eta}_{tt})-m_{b}f_{bb}(\eta _{bb}-\bar{\eta}_{bb})), 
\]
and $\phi ^{-}=\bar{\phi}^{+}$. This result identifies the bosons $\phi
^{\pm }$ and $\phi ^{0}$. Then we search for constants $f_{W}$ and $f_{Z}$
such that these Goldstone bosons disappear from the difference $\Gamma
_{\eta \eta }^{\prime }=\Gamma _{\eta \eta }-\Gamma _{\phi \phi }$, where

\[
\Gamma _{\phi \phi }=\frac{N_{c}}{f_{W}}(\partial _{\mu }\phi ^{+})(\partial
^{\mu }\phi ^{-})+\frac{N_{c}}{2f_{Z}}(\partial _{\mu }\phi ^{0})(\partial
^{\mu }\phi ^{0}). 
\]
We find 
\[
f_{W}=m_{t}^{2}f_{tb}^{\prime }+m_{b}^{2}f_{bt}^{\prime },\qquad f_{Z}=\frac{%
1}{2}(m_{t}^{2}f_{tt}+m_{b}^{2}f_{bb}). 
\]
Next, we determine the linearized gauge transformations of the Goldstone
bosons, demanding that $\Gamma _{\phi \phi }+\Gamma _{A\eta }$ be invariant
up to $\mathcal{O}(A)$, where $A$ denotes a generic gauge field. We find 
\begin{equation}
\delta W_{\mu }^{\pm }=\partial _{\mu }C^{\pm },\qquad \delta Z_{\mu
}=\partial _{\mu }C^{0},\qquad \delta \phi ^{\pm }=gf_{W}C^{\pm },\qquad
\delta \phi ^{0}=\tilde{g}f_{Z}C^{0}.  \label{ga}
\end{equation}
Finally, we add $A^{2}$-terms to have gauge invariance at the linearized
level. In total, the relevant quadratic contributions $\Delta _{2}\Gamma $
to the $\Gamma $ functional read 
\[
\Delta _{2}\Gamma =\Gamma _{\eta \eta }^{\prime }+\frac{N_{c}}{f_{W}}%
(\partial _{\mu }\phi ^{+}-gf_{W}W_{\mu }^{+})(\partial ^{\mu }\phi
^{-}-gf_{W}W^{\mu -})+\frac{N_{c}}{2f_{Z}}(\partial _{\mu }\phi ^{0}-\tilde{g%
}f_{Z}Z_{\mu })(\partial ^{\mu }\phi ^{0}-\tilde{g}f_{Z}Z^{\mu }). 
\]
Choosing the unitary gauge-fixing $\phi ^{\pm }=\phi ^{0}=0$, we can read
the gauge-boson squared masses 
\begin{equation}
m_{W}^{2}=N_{c}g^{2}f_{W}\sim \frac{N_{c}g^{2}}{32\pi ^{2}}m_{t}^{2}\ln 
\frac{\Lambda _{L}^{2}}{m_{t}^{2}},\qquad m_{Z}^{2}=N_{c}\tilde{g}%
^{2}f_{Z}\sim \frac{\tilde{g}^{2}}{g^{2}}m_{W}^{2}.  \label{auto}
\end{equation}
The first formula can be converted into a relation between the Fermi
constant and the mass of the top quark, namely 
\begin{equation}
\frac{1}{G_{F}}=\frac{N_{c}m_{t}^{2}}{4\pi ^{2}\sqrt{2}}\ln \frac{\Lambda
_{L}^{2}}{m_{t}^{2}}.  \label{mast}
\end{equation}
Using our estimated value $\Lambda _{L}=10^{14}\mathrm{GeV}$, we find $%
m_{t}=171.6\mathrm{GeV}$. This ``too-good'' agreement has no simple
explanation, as far as we know, also taking into account that from a
quantitative point of view our rough large $N_{c}$ approximation contains a
good 50\% margin of error\footnote{%
Call ``1'' the leading order of the large $N_{c}$ expansion. Resumming
powers of 1/3 from 1 to infinity we get 1/2, so, generically speaking, a
``1'' could be anything between 1/2 and 3/2.}.

The quantities $f_{W}$ and $f_{Z}$ are related in a non-straightforward way.
We find 
\[
\rho \equiv \frac{\tilde{g}^{2}m_{W}^{2}}{g^{2}m_{Z}^{2}}=\frac{f_{W}}{f_{Z}}%
=2\frac{m_{t}^{2}f_{tb}^{\prime }+m_{b}^{2}f_{bt}^{\prime }}{%
m_{t}^{2}f_{tt}+m_{b}^{2}f_{bb}}. 
\]
The explicit computation for $\Lambda _{L}\gg m_{t}\gg m_{b}$ shows that the
values of $f_{W}$ and $f_{Z}$ are actually close. Indeed, the relation $\rho
\sim 1$ is fulfilled not only when there is an approximate custodial
symmetry (which would imply approximately equal quark masses), but also in
the opposite situation, namely when one quark mass is much larger than the
other one. In this case the deviations from $\rho =1$ are, at low energies,
just those predicted by the usual Standard Model results, as already noted
in \cite{tcond2}.

The vertices of the effective action can be derived calculating diagrams
with one fermion loop and more external $A$ and $\eta $ legs. Once the gap
equations (\ref{gap2}) are used, every other contribution is convergent and
unambiguously determined. In principle, using the large $N_{c}$ expansion we
can calculate the effective action with the desired precision.

We have shown that a forth Goldstone boson is associated with the breaking
of $U(1)_{A}$. This boson becomes massive because of the $U(1)_{A}$-anomaly.
On the other hand, quark masses, which are not due to an explicit symmetry
breaking, do not contribute to the mass of this boson.

\paragraph{Composite Higgs bosons}

So far, we have assumed that all $\eta $-entries were independent, which
amounts to have two independent composite Higgs doublets. The situation with
a single composite Higgs doublet can be retrieved choosing 
\begin{equation}
\tau =\frac{m_{t}}{v}\sqrt{2}\left( 
\begin{tabular}{cc}
$H_{2}$ & $-H_{1}$ \\ 
$\kappa \bar{H}_{1}$ & $\kappa \bar{H}_{2}$%
\end{tabular}
\right) ,\qquad \kappa =\frac{m_{b}}{m_{t}},  \label{hig}
\end{equation}
where $v/\sqrt{2}$ is the Higgs vev. Substituting (\ref{hig}) in $\Gamma
_{\tau \tau }$ we find the three Goldstone bosons associated with the
breaking of $SU(2)_{L}\times U(1)_{Y}$, plus one neutral massive scalar with
squared mass 
\[
4\frac{m_{t}^{4}f_{tt}+m_{b}^{4}f_{bb}}{m_{t}^{2}f_{tt}+m_{b}^{2}f_{bb}}\sim
4m_{t}^{2}, 
\]
to be identified with the composite Higgs scalar. Its mass is $2m_{t}$, as
originally suggested by Nambu \cite{nambu}. This value is far from the
expected Higgs mass, but taking into account of our 50\% margin of error,
the final, exact formula could still give $m_{H}\sim m_{t}$, which would be
compatible with present expectations.

More generally, we can introduce three doublets for each family (if there
exist no right-handed neutrinos): two for the quarks and one for the
leptons. Actually, to allow for mixing among families we can just take the
Yukawa vertices $\mathcal{L}_{Y}$ and promote every product $YH$ to an
independent field $\tau $: 
\[
\mathcal{L}_{\tau \psi \psi }+\mathcal{L}_{\tau \tau
}=-\sum_{a,b=1}^{3}\left( \bar{L}^{a\hspace{0.01in}i}\tau _{\ell
}^{ab,i}\ell _{R}^{b}+\bar{Q}_{R}^{ai}\tau _{q}^{ab,ij}Q_{L}^{bj}+\text{h.c.}%
\right) +V_{\ell q}(\tau _{\ell },\tau _{q}), 
\]
where $V_{2}$ is the most general quadratic polynomial compatible with the
symmetries of the theory. To generate Majorana masses for neutrinos we need
extra four fermion vertices encoded in \cite{linde} 
\[
\mathcal{L}_{\tau LL}^{\prime }+\mathcal{L}_{\tau \tau }^{\prime
}=-\sum_{a,b=1}^{3}(\bar{L}^{c})^{a\hspace{0.01in}i}\varepsilon ^{ij}\tau
_{L}^{ab,jk}L^{bk}+\text{h.c.}+V_{\ell \ell }^{\prime }(\tau _{L}). 
\]
We can add analogous Lorentz violating terms containing fields $\rho $, but
we know that we can neglect such terms both for the search of the minimum of
the potential and to derive the induced low-energy effective action.

Note that in the lepton sector we have no analogue of the large $N_{c}$
expansion to justify our arguments. We still expect, however, that the
minimum exists and the dynamical symmetry breaking takes place.

\section{Goldstone theorem in Lorentz violating theories}

\setcounter{equation}{0}

In the previous section we have used the large $N_{c}$ expansion to prove
the dynamical symmetry breaking and study bound states, among which the
Goldstone bosons. However, the Goldstone theorem is an exact result, that
can be derived without making use of expansions or approximations. In this
section we show how to generalize it to Lorentz violating theories. We
assume that the theory becomes Lorentz invariant at large distances. We do
not need to assume that the scale of symmetry breaking is much smaller than
the scale of Lorentz violation.

Let $\omega (x)$ denote a generic field operator and $J_{\mu }=(J^{\hat{\mu}%
},J^{\bar{\mu}})$ the conserved current associated with a continuous global
symmetry. Current conservation $\partial _{\mu }J^{\mu }=\partial _{\hat{\mu}%
}J^{\hat{\mu}}+\partial _{\bar{\mu}}J^{\bar{\mu}}=0$ still implies 
\begin{equation}
\frac{\mathrm{d}}{\mathrm{d}t}[Q(t),\omega (0)]=0,\qquad Q(t)=\int \mathrm{d}%
^{3}x\hspace{0.01in}J^{0}(t,\mathbf{x}).  \label{commo}
\end{equation}
Indeed, the last term of the equality 
\[
0=\int \mathrm{d}^{3}x\hspace{0.01in}[\partial _{\mu }J^{\mu }(x),\omega (0)%
]=\frac{\mathrm{d}}{\mathrm{d}t}[Q(t),\omega (0)]+\int_{S^{\infty }}\mathrm{d%
}\mathbf{s}\cdot \hspace{0.01in}[\mathbf{J}(t,\mathbf{x}),\omega (0)] 
\]
is equal to zero, because for \textit{large} space-like separations the
commutator $[\mathbf{J}(t,\mathbf{x}),\omega (0)]$ vanishes. This property
holds also in our Lorentz violating theories, since they become Lorentz
invariant at large distances, if the vacuum is Lorentz invariant. For
generic space-like separations a commutator does not need to vanish.

The symmetry is spontaneously broken if the commutator $[Q(t),\omega (0)]$
has a non-vanishing expectation value $u$. Then, inserting a complete set of
intermediate states $|n\rangle $ and using translational invariance we have 
\[
u=\sum_{n}(2\pi )^{3}\delta ^{(3)}(\mathbf{p}_{n})\hspace{0.01in}\left[ 
\mathrm{e}^{-iE_{n}t}\langle 0|J_{0}(0)|n\rangle \langle n|\omega
(0)|0\rangle -\mathrm{e}^{iE_{n}t}\langle 0|\omega (0)|n\rangle \langle
n|J_{0}(0)|0\rangle \right] . 
\]
Since $u$ is constant, because of (\ref{commo}), there must exist a state $|%
\bar{n}\rangle $ such that $E_{\bar{n}}=\mathbf{p}_{\bar{n}}=0$ and 
\[
\langle 0|J_{0}(0)|\bar{n}\rangle \neq 0,\qquad \langle \bar{n}|\omega
(0)|0\rangle \neq 0. 
\]

Now, consider the case of $SU(2)_{L}\times U(1)_{Y}$ spontaneously broken to 
$U(1)_{Q}$. We have composite fields $\omega ^{\pm ,0}$ and Goldstone bosons 
$\phi ^{\pm ,0}$, such that 
\begin{equation}
\langle 0|\omega ^{\pm ,0}(0)|\phi ^{\pm ,0}\rangle \neq 0,\qquad \langle
0|J_{0}^{\pm ,0}(0)|\phi ^{\pm ,0}\rangle \neq 0,  \label{ar}
\end{equation}
where $J_{\mu }^{\pm ,0}$ are the currents associated with the broken
generators. The form of $\langle 0|J_{0}^{\pm ,0}(0)|\phi ^{\pm ,0}\rangle $
is no longer constrained by Lorentz invariance. Instead, we have, in
momentum space, 
\begin{equation}
\langle 0|J_{\mu }^{\pm ,0}(0)|\phi ^{\pm ,0}(p)\rangle =if_{\pm ,0}(\hat{p}%
_{\mu }+\zeta _{\pm ,0}\bar{p}_{\mu }),  \label{br}
\end{equation}
where $f_{\pm ,0}$ and $\zeta _{\pm ,0}$ may depend on $\bar{p}^{2}$.
Current conservation implies $\hat{p}^{2}-\zeta _{\pm ,0}\bar{p}^{2}=0$ on
shell, which determines the $\phi $-kinetic terms. It also eliminates the $%
\hat{p}^{2}$-dependence and ensures that $\zeta _{\pm ,0}$ are real. The
conservation of electric charge implies $f_{+}=f_{-}^{*}$, $\zeta _{+}=\zeta
_{-}\equiv \zeta $ and that $f_{0}$ is also real. The effective lagrangian
incorporating such pieces of information reads in the quadratic
approximation 
\begin{eqnarray*}
\mathcal{L}_{\text{eff}} &=&(\hat{\partial}_{\mu }\phi ^{+}-g\hat{W}_{\mu
}^{+}f_{+})(\hat{\partial}_{\mu }\phi ^{-}-gf_{-}\hat{W}_{\mu }^{-})-(\bar{%
\partial}_{\mu }\phi ^{+}-g\bar{W}_{\mu }^{+}f_{+})\zeta (\bar{\partial}%
_{\mu }\phi ^{-}-gf_{-}\bar{W}_{\mu }^{-}) \\
&&+\frac{1}{2}(\hat{\partial}_{\mu }\phi ^{0}-\tilde{g}\hat{Z}_{\mu
}f_{0})^{2}-\frac{1}{2}(\bar{\partial}_{\mu }\phi ^{0}-\tilde{g}\bar{Z}_{\mu
}f_{0})\zeta _{0}(\bar{\partial}_{\mu }\phi ^{0}-\tilde{g}f_{0}\bar{Z}_{\mu
}).
\end{eqnarray*}
We have the linearized gauge symmetry 
\begin{equation}
\delta W_{\mu }^{\pm }=\partial _{\mu }C^{\pm },\qquad \delta Z_{\mu
}=\partial _{\mu }C^{0},\qquad \delta \phi ^{\pm }=gf_{\pm }C^{\pm },\qquad
\delta \phi ^{0}=\tilde{g}f_{0}C^{0}.  \label{gs}
\end{equation}
Choosing the gauge-fixing $\phi ^{\pm ,0}=0$, we find the $W$ and $Z$ mass
terms 
\[
\mathcal{L}_{\text{m}}=g^{2}(\hat{W}_{\mu }^{+}f_{+}f_{-}\hat{W}_{\mu }^{-}-%
\bar{W}_{\mu }^{+}f_{+}\zeta f_{-}\bar{W}_{\mu }^{-})+\frac{1}{2}\tilde{g}%
^{2}(\hat{Z}_{\mu }f_{0}^{2}\hat{Z}_{\mu }-\bar{Z}_{\mu }f_{0}\zeta _{0}f_{0}%
\bar{Z}_{\mu }), 
\]
as in the Proca version of Lorentz violating gauge theories \cite
{LVgauge1suA}.

Observe that at this level, our construction is unable to relate the $W$ and 
$Z$ masses. At low energies, when Lorentz symmetry is restored ($\zeta
=\zeta _{0}=1$) and $f_{\pm }$, $f_{0}$ can be taken to be constant, we have 
\begin{equation}
\mathcal{L}_{\text{m}}=g^{2}f_{+}f_{-}W_{\mu }^{+}W^{-\mu }+\frac{1}{2}%
\tilde{g}^{2}f_{0}^{2}Z_{\mu }Z^{\mu },  \label{br2}
\end{equation}
so on general grounds we are unable to predict $\rho =1$, actually 
\begin{equation}
\rho =\frac{f_{+}f_{-}}{f_{0}^{2}},  \label{w}
\end{equation}
In the Standard Model, $\rho =1$ follows from $f_{+}f_{-}=$ $%
f_{0}^{2}=v^{2}/4$. In the model of the previous section, it follows from
explicit calculation in the large $N_{c}$ expansion, for quarks with very
different masses.

When there exists a symmetry $SU(2)_{R}$, for example a ``custodial''
symmetry \cite{sikivie}, the right-handed quarks can be collected into $%
SU(2)_{R}$ doublets, and the quark sector becomes $SU(2)_{L}\times SU(2)_{R}$%
-invariant. If the vacuum preserves $SU(2)_{L+R}$, (\ref{br}) is modified to 
\begin{equation}
\langle 0|J_{\mu }^{\pm ,0}(0)|\phi ^{\pm ,0}(p)\rangle =i\tilde{f}(\hat{p}%
_{\mu }+\tilde{\zeta}\bar{p}_{\mu }),  \label{bra1}
\end{equation}
with unique $\tilde{f}$ and $\tilde{\zeta}$ for $W^{\pm }$ and $Z$. Then $%
\rho =1$ follows (at the tree level).

\section{Conclusions}

\setcounter{equation}{0}

The Standard-Extended Model (\ref{simplL}) proposed in ref. \cite{LVSM}, and
its variants, such as (\ref{simplL2}), contain interactions that are
non-renormalizable by ordinary power counting, but are renormalizable by
weighted power counting, thanks to the high-energy Lorentz violation. An
interesting feature of those models is that they become very simple at high
energies ($\gtrsim 10^{14}$GeV), because all gauge and Higgs interactions,
being super-renormalizable, disappear. There survives a four fermion model
in two weighted dimensions, which admits a dynamical symmetry breaking. In
spite of the fact that such a four fermion model is Lorentz violating, the
dynamically generated vacuum and the low energy effective action are Lorentz
invariant. We have therefore focused our attention on the scalarless variant (%
\ref{noH}) of the model of \cite{LVSM}, which contains no elementary scalar
field. In the large $N_{c}$ expansion we have seen that the dynamical
symmetry breaking generates composite massive Higgs bosons and gives masses
to fermions and gauge bosons. The model is predictive, in the sense that it
does not contain the ambiguities of previous approaches, which relied on the
non-renormalizable Nambu--Jona-Lasinio mechanism, and is candidate to explain the observed low energy physics. The leading order of the
large $N_{c}$ expansion, with gauge interactions switched off, does not
allow us to make very precise quantitative predictions, although the
relation (\ref{mast}) between the Fermi constant and the top mass turns out
to be mysteriously right.

A step forward towards more precise predictions is to include the effects of
the RG flow from energies $\sim m_{t}$ to $\Lambda _{L}$, and study the
condition of compositeness at energies $\sim \Lambda _{L}$ \cite{tcond2}.
However, in our Lorentz violating theories the RG flow is considerably
different from the usual one: it coincides with the usual one at energies $%
\sim m_{t}$, since the low energy theory (with composite Higgs bosons
included) is renormalizable by ordinary power counting; on the other hand,
it changes completely as we move to energies $\sim \Lambda _{L}$, because
there, gauge interactions do not run. Work is in progress in this direction.

What we have done in this paper is not only to describe low energy effects
of high energy Lorentz violations, but also show that they can be consistent
with low energy Lorentz invariance. This fact was not obvious a priori.

Our mechanism can of course take place also in the Higgsed models (\ref
{simplL}) and (\ref{simplL2}), if the four fermion vertices are chosen
appropriately. There, its effects sum to those of the elementary Higgs
doublet. It can also be applied to Standard Model extensions that contain
new types of fermions, interacting by four fermion vertices, such as those
considered in ref.s \cite{cynolter}.

\vskip 20truept \noindent {\Large \textbf{Acknowledgments}}

\vskip 10truept

I am grateful to E. Ciuffoli, J. Evslin, E. Gabrielli, E. Guadagnini, S.B.
Gudnason, E. Meggiolaro, M. Mintchev, G. Paffuti, S. Rychkov and A. Strumia
for helpful discussions. I also thank J. Evslin for pointing out relevant
references.

\vskip 20truept \noindent {\Large \textbf{Appendix A: Simplification of the
pure gauge sector}}

\vskip 10truept

\renewcommand{\theequation}{A.\arabic{equation}} \setcounter{equation}{0}

In this appendix we describe a weight reassignment that is useful to
simplify the gauge sector. Indeed, the quadratic gauge field lagrangian $%
\mathcal{L}_{Q}$ generates an involved propagator \cite{LVgauge1suA}. Recall
that in Lorentz violating gauge theories the BRST\ symmetry is the same as
usual, 
\begin{eqnarray*}
sA_{\mu }^{a} &=&D_{\mu }^{ab}C^{b}=\partial _{\mu }C^{a}+gf^{abc}A_{\mu
}^{b}C^{c},\qquad sC^{a}=-\frac{g}{2}f^{abc}C^{b}C^{c}, \\
s\bar{C}^{a} &=&B^{a},\qquad sB^{a}=0,\qquad s\psi
^{i}=-gT_{ij}^{a}C^{a}\psi ^{j},
\end{eqnarray*}
etc., where $B^{a}$ are Lagrange multipliers for the gauge-fixing. The most
convenient gauge-fixing is 
\begin{equation}
\mathcal{L}_{\text{gf}}=s\Psi ,\qquad \Psi =\bar{C}^{a}\left( -\frac{\lambda 
}{2}B^{a}+\mathcal{G}^{a}\right) ,\qquad \mathcal{G}^{a}\equiv \hat{\partial}%
\cdot \hat{A}^{a}+\zeta \left( \bar{\upsilon}\right) \bar{\partial}\cdot 
\bar{A}^{a},  \label{gf}
\end{equation}
where $\lambda $ is a dimensionless, weightless constant, $\bar{\upsilon}%
\equiv -\bar{\partial}^{2}/\Lambda _{L}^{2}$ and $\zeta $ is a polynomial of
degree $n-1$. The total gauge-fixed action is 
\begin{equation}
\mathcal{S}=\int \mathrm{d}^{d}x\left( \mathcal{L}_{Q}+\mathcal{L}_{I}+%
\mathcal{L}_{\text{gf}}\right) ,  \label{basis}
\end{equation}
where $\mathcal{L}_{I}$ collects the terms that are at least cubic in the
field strength.

The gauge-field propagator can be worked out from the free subsector of (\ref
{basis}), after integrating $B^{a}$ out, which amounts to add $(\mathcal{G}%
^{a})^{2}/(2\lambda )$ to the quadratic lagrangian $\mathcal{L}_{Q}$%
\footnote{%
With respect to the $\mathcal{L}_{Q}$ of (\ref{varie}), the most general
quadratic gauge field lagrangian \cite{LVgauge1suA,LVgauge1suAbar} can
contain another polynomial $\xi (\Upsilon )$. Here we set it to zero, since
the weight reassignment excludes it anyway from the theory with simplified
gauge sector.}. We find, in Euclidean space, 
\begin{equation}
\langle A(k)\ A(-k)\rangle =\left( 
\begin{array}{cc}
\langle \hat{A}\hat{A}\rangle & \langle \hat{A}\bar{A}\rangle \\ 
\langle \bar{A}\hat{A}\rangle & \langle \bar{A}\bar{A}\rangle
\end{array}
\right) =\left( 
\begin{array}{cc}
u & r\hat{k}\bar{k} \\ 
r\bar{k}\hat{k} & \quad v\bar{\delta}+t\bar{k}\bar{k}
\end{array}
\right) ,  \label{pros}
\end{equation}
where 
\[
u=\frac{\lambda \hat{k}^{2}+\frac{\zeta ^{2}}{\eta }\bar{k}^{2}}{%
D^{2}(1,\zeta )},\qquad r=\frac{\lambda -\frac{\zeta }{\eta }}{D^{2}(1,\zeta
)},\qquad v=\frac{1}{D(\eta ,\tau )},\qquad t=\frac{\left( \eta \lambda +%
\frac{\tau }{\eta }-2\zeta \right) \hat{k}^{2}+\left( \tau \lambda -\zeta
^{2}\right) \bar{k}^{2}}{D(\eta ,\tau )D^{2}(1,\zeta )}, 
\]
and $D(x,y)\equiv x\hat{k}^{2}+y\bar{k}^{2}$. Now $\eta $, $\tau $ and $%
\zeta $, as well as $x$ and $y$, are functions of $\bar{k}^{2}/\Lambda
_{L}^{2}$. The ghost propagator is $1/D(1,\zeta )$.

If $\eta \neq 1$ the propagator is not regular \cite
{LVgauge1suA,LVgauge1suAbar}, because for $\hat{k}$ large $\langle \bar{A}%
\bar{A}\rangle $ behaves like $1/(\eta \hat{k}^{2})$ and $\eta $ depends
only on $\bar{k}^{2}$. Thus the $\hat{k}$-integrals may contain ``spurious''
subdivergences. A separate power-counting analysis is necessary to show that
such subdivergences are absent under certain conditions \cite{LVgauge1suAbar}%
, fulfilled by the model (\ref{simplL}).

An alternative solution, which avoids the problem from the start, is to set $%
\eta =1$, in which case the propagator clearly becomes regular. We show that
this choice is consistent with renormalization, if combined with a
rearrangement of the weight assignments and other choices, such as
restricting $\tau $ to be a polynomial of degree $n-1$ (which we then denote
by $\tau ^{\prime }$). The new quadratic gauge field lagrangian reads 
\begin{equation}
\mathcal{L}_{Q}^{\prime }=\frac{1}{2}F_{\hat{\mu}\bar{\nu}}^{2}-\frac{1}{4}%
F_{\bar{\mu}\bar{\nu}}\tau ^{\prime }(\bar{\Upsilon})F_{\bar{\mu}\bar{\nu}}
\label{lq}
\end{equation}
and admits a nice diagonal propagator 
\begin{equation}
\left( 
\begin{array}{cc}
\langle \hat{A}\hat{A}\rangle & \langle \hat{A}\bar{A}\rangle \\ 
\langle \bar{A}\hat{A}\rangle & \langle \bar{A}\bar{A}\rangle
\end{array}
\right) =\frac{1}{D(1,\tau ^{\prime })}\left( 
\begin{array}{cc}
\tau ^{\prime } & 0 \\ 
0 & \bar{\delta}
\end{array}
\right) ,  \label{pd}
\end{equation}
in the ``Feynman'' gauge 
\begin{equation}
\zeta =\lambda =\tau ^{\prime }.  \label{fey}
\end{equation}
After integrating $B$ out, the propagator (\ref{pd}) is obtained adding 
\begin{equation}
\frac{1}{2}(\mathcal{G}^{a})\frac{1}{\tau ^{\prime }}(\mathcal{G}^{a})
\label{tp}
\end{equation}
to $\mathcal{L}_{Q}^{\prime }$. Note that (\ref{tp}) is non-local, because
the gauge condition $\lambda =\tau ^{\prime }$ contained in (\ref{fey})
implies that the constant $\lambda $ is replaced with a function of $\bar{k}%
^{2}/\Lambda _{L}^{2}$. This replacement is legitimate, since the action (%
\ref{basis}) is local before integrating $B$ out, and $B$ is non-propagating
(see (\ref{gf})).

The consistency of (\ref{lq}) is explained by a simple weight rearrangement,
with gauge field components acquiring higher weights and the gauge coupling
acquiring a lower weight, such that the product $gA$ maintains the same
weight. Denoting weights with square brakets, we have, by covariance, $[g%
\hat{A}]=[\hat{\partial}]=1$ and $[g\bar{A}]=[\bar{\partial}]=1/n$. The
field strength is split into $\tilde{F}_{\mu \nu }\equiv F_{\hat{\mu}\bar{\nu%
}}$ and $\bar{F}_{\mu \nu }\equiv F_{\bar{\mu}\bar{\nu}}$. The kinetic
lagrangian $\mathcal{L}_{Q}^{\prime }$ contains $\tilde{F}^{2}$, so $\tilde{F%
}$ must have weight \dj $/2$. Since $[\tilde{F}]=[\bar{\partial}]+[\hat{A}]=[%
\hat{\partial}]+[\bar{A}]$, we have 
\begin{equation}
\lbrack \hat{A}]=\frac{\text{\dj }}{2}-\frac{1}{n},\qquad [\bar{A}]=\frac{%
\text{\dj }}{2}-1,\qquad [\tilde{F}]=\frac{\text{\dj }}{2},\qquad [\bar{F}]=%
\frac{\text{\dj }}{2}-1+\frac{1}{n}.  \label{w1}
\end{equation}
The weight of the gauge coupling is 
\begin{equation}
\lbrack g]=1+\frac{1}{n}-\frac{\text{\dj }}{2}.  \label{w2}
\end{equation}
Observe that $[g]>0$ in four dimensions, for $n>1$, where gauge interactions
are always super-renormalizable. We also find $[\zeta ]=[\lambda ]=[\tau
^{\prime }]=2-2/n$, which implies that $\tau ^{\prime }$ must be of order $%
n-1$ and makes the gauge choice (\ref{fey}) renormalizable.

The quadratic terms of the ghost Lagrangian contain $\bar{C}\hat{\partial}%
^{2}C$ and $\lambda B^{2}$, which have weight \dj , so we have the weight
assignments 
\begin{equation}
\lbrack C]=[\bar{C}]=\frac{\text{\dj }}{2}-1,\qquad [s]=\frac{1}{n},\qquad
[B]=\frac{\text{\dj }}{2}-1+\frac{1}{n}.  \label{w3}
\end{equation}

In our case (\dj $=2$, $n=3$) we have $[g]=1/3$. By covariance, the coupling 
$\bar{g}$ attached to the scalar legs must satisfy $[\bar{g}]\leq [g]$ \cite
{LVgauge1suAbar}, so we lower $[\bar{g}]$ from $1/2$ to $1/3$. All other
weights are unchanged. Then the most general lagrangian is (\ref{simplL2})
plus the extra terms 
\begin{eqnarray}
&&\bar{g}^{6}H^{8},\quad \bar{g}^{4}\bar{D}^{2}H^{6},\quad \bar{g}^{2}\bar{D}%
^{4}H^{4},\quad g\bar{g}^{2}\bar{D}^{2}\bar{F}H^{4},\quad g\bar{D}^{4}\bar{F}%
H^{2},\quad g^{2}\bar{g}^{2}\bar{F}^{2}H^{4},\quad g^{2}\bar{D}^{2}\bar{F}%
^{2}H^{2},\quad g^{3}\bar{F}^{3}H^{2},  \nonumber \\
&&g^{2}\bar{F}^{4},\qquad g\bar{D}^{2}\bar{F}^{3},\qquad g\bar{\varepsilon}%
\tilde{F}\bar{D}^{2}H^{2},\qquad g^{2}\bar{\varepsilon}\tilde{F}\bar{F}%
H^{2},\qquad \bar{\varepsilon}\tilde{F}\bar{D}^{2}\bar{F},\qquad g\bar{%
\varepsilon}\bar{F}\tilde{F}\bar{F},\qquad \bar{\varepsilon}\bar{F}\hat{D}%
\bar{D}\bar{F},  \nonumber \\
&&g\bar{g}\bar{\psi}\psi \bar{F}H,\qquad \bar{g}\bar{\psi}\psi \bar{D}%
^{2}H,\qquad \bar{g}^{2}\bar{\psi}\psi \bar{D}H^{2},\qquad \bar{g}^{3}\bar{%
\psi}\psi H^{3},\qquad  \label{tu}
\end{eqnarray}
and those obtained suppressing some fields and/or derivatives, where $\bar{%
\varepsilon}$ is the $\varepsilon $-tensor with three space indices. The
extra terms (\ref{tu}) can be consistently dropped, because they are not
generated back by renormalization.

The two models (\ref{simplL}) and (\ref{simplL2}) correspond to the basic
weight assignments $[s]=1$ and $[s]=1/n$, respectively. All intermediate
situations $[s]=k/n$, $k=1,\ldots n$ are actually allowed, with suitable
weight reassignments. Observe that the construction of this Appendix is
possible only because spacetime is broken into space and time. Indeed, other
types of breakings are disfavored \cite{LVgauge1suA,LVgauge1suAbar}.

\vskip 20truept \noindent {\Large \textbf{Appendix B: Useful mathematical
formulas}}

\vskip 10truept

\renewcommand{\theequation}{B.\arabic{equation}} \setcounter{equation}{0}

In this appendix we collect a number of mathematical formulas and relations
that are used in the paper.

Define the function 
\begin{equation}
v(r)=-\int^{\Lambda /\Lambda _{L}}\frac{\text{\textrm{d}}^{4}p}{(2\pi )^{4}}%
\ln (p^{\prime \prime \hspace{0.01in}2}+r),  \label{d1}
\end{equation}
where $r>0$, the integral is in Euclidean space, 
\begin{equation}
\hat{p}^{\prime \prime \hspace{0.01in}\mu }=\hat{p}^{\mu },\qquad \bar{p}%
^{\prime \prime \hspace{0.01in}\mu }=\bar{p}^{\mu }(\bar{p}^{2}+1)
\label{uy2}
\end{equation}
and $\Lambda $ is a UV cut-off. We want to study the Taylor expansion 
\begin{equation}
v(r)=v(r_{0})+(r-r_{0})v^{\prime }(r_{0})+\frac{1}{2}(r-r_{0})^{2}v^{\prime
\prime }(r_{0})+\mathcal{O}((r-r_{0})^{3})  \label{vt}
\end{equation}
of this function in the neighborhood of a generic point $r_{0}$. Observe
that the second derivative 
\begin{equation}
v^{\prime \prime }(r_{0})=\int \frac{\text{\textrm{d}}^{4}p}{(2\pi )^{4}}%
\frac{1}{(p^{\prime \prime \hspace{0.01in}2}+r_{0})^{2}}  \label{2d}
\end{equation}
is convergent and strictly positive. On the other hand, the first derivative
is logarithmically divergent. By weighted power counting, its divergent part
is independent of $r_{0}$. We have 
\begin{equation}
\frac{v^{\prime }(r_{0})-v^{\prime }(r_{1})}{r_{0}-r_{1}}=\int \frac{\text{%
\textrm{d}}^{4}p}{(2\pi )^{4}}\frac{1}{(p^{\prime \prime \hspace{0.01in}%
2}+r_{0})(p^{\prime \prime \hspace{0.01in}2}+r_{1})}>0.  \label{rinc}
\end{equation}
Adding and subtracting $1/(\hat{p}^{\hspace{0.01in}2}+(\bar{p}%
^{2})^{3}+r_{0})$ to the integrand of $v^{\prime }(r_{0})$ we can also write 
\begin{equation}
v^{\prime }(r_{0})=\bar{v}^{\prime }(r_{0})-\frac{1}{12\pi ^{2}}\ln \frac{%
2\Lambda ^{3}}{\Lambda _{L}^{3}\sqrt{r_{0}}},  \label{g3}
\end{equation}
up to $\mathcal{O}(1/\Lambda )$, where 
\begin{equation}
\bar{v}^{\prime }(r_{0})=\int \frac{\text{\textrm{d}}^{4}p}{(2\pi )^{4}}%
\frac{\bar{p}^{2}(2\bar{p}^{2}+1)}{(p^{\prime \prime \hspace{0.01in}%
2}+r_{0})(\hat{p}^{\hspace{0.01in}2}+(\bar{p}^{2})^{3}+r_{0})}  \label{g2}
\end{equation}
is finite and positive.

Now we approximate the expansion (\ref{vt}) for $r_{0}\ll 1$. To study the
right-hand side of (\ref{2d}) we note that the integral diverges
logarithmically for small $r_{0}$, so it is sufficient to look for the
corresponding logarithm. We find 
\begin{equation}
v^{\prime \prime }(r_{0})\sim -\frac{1}{16\pi ^{2}}\ln r_{0},\qquad \text{%
for }r_{0}\ll 1.  \label{v2}
\end{equation}
Integrating this expression, we also find 
\begin{equation}
v^{\prime }(r_{0})=-\int^{\Lambda /\Lambda _{L}}\frac{\text{\textrm{d}}^{4}p%
}{(2\pi )^{4}}\frac{1}{p^{\prime \prime \hspace{0.01in}2}+r_{0}}\sim -\frac{1%
}{16\pi ^{2}}\left( 2\ln \frac{\Lambda ^{2}}{\Lambda _{L}^{2}}+r_{0}\ln
r_{0}\right) .  \label{g1}
\end{equation}
up to $\mathcal{O}(1/\Lambda )$. The arbitrary constant can be determined
calculating $v^{\prime }(0)$.

Another useful integral is 
\begin{equation}
I_{1}=\int \frac{\text{\textrm{d}}^{4}p}{(2\pi )^{4}}\frac{\hat{p}^{2}-\bar{p%
}^{\prime \prime \hspace{0.01in}2}}{(p^{\prime \prime \hspace{0.01in}%
2}+r_{0})^{2}}=r_{0}v^{\prime \prime }(r_{0})>0.  \label{an1}
\end{equation}
This formula is proved using the identity 
\begin{equation}
\int_{-\infty }^{+\infty }\frac{\text{\textrm{d}}\hat{p}}{2\pi }\frac{\hat{p}%
^{2}-a_{k}x}{(\hat{p}^{\hspace{0.01in}2}+x)^{k}}=0,\qquad a_{k}=\frac{\Gamma
\left( k-\frac{3}{2}\right) }{2\hspace{0.01in}\Gamma \left( k-\frac{1}{2}%
\right) },\qquad x>0,\qquad k>\frac{3}{2},  \label{ant}
\end{equation}
for $k=2$.

Next, define the functions 
\begin{equation}
\left( f_{ij},f_{ij}^{\prime },f_{ij}^{\prime \prime }\right) (p^{2})=\frac{1%
}{(4\pi )^{2}}\int_{0}^{1}\mathrm{d}x\hspace{0.01in}\left( 1,x,x(1-x)\right)
\left[ \ln \frac{\Lambda _{L}^{2}}{m_{i}^{2}x+m_{j}^{2}(1-x)-p^{2}x(1-x)}%
-1\right] ,  \label{ff}
\end{equation}
where $i,j$ can have the values $t$ and $b$. While $f_{ij}$ is clearly
symmetric, $f_{ij}^{\prime }$ satisfies 
\[
f_{ij}^{\prime }+f_{ji}^{\prime }=f_{ij}. 
\]
In the range $0\leqslant p\leqslant m_{t}$ the functions (\ref{ff}) do not
depend on $p$ very much and can be treated as constants, calculated for $p=0$%
. Using $m_{b}\ll m_{t}\ll \Lambda _{L}$, we have 
\begin{equation}
f_{ii}\sim \frac{1}{(4\pi )^{2}}\ln \frac{\Lambda _{L}^{2}}{m_{i}^{2}}%
,\qquad f_{tb}=f_{bt}\sim f_{tt},\qquad f_{ij}^{\prime }\sim \frac{1}{2}%
f_{ij},\qquad f_{ij}^{\prime \prime }\sim \frac{1}{6}f_{ij}.  \label{appro}
\end{equation}

\end{document}